\documentclass[aps,prd,preprintnumbers,superscriptaddress,nofootinbib,showpacs,twocolumn]{revtex4}%
\usepackage[dvipdfmx]{graphicx}
\usepackage{bm,latexsym,amsmath,amssymb,amsfonts,mathrsfs}
\usepackage{color}
\input{colordvi.tex}
\allowdisplaybreaks[1]
\usepackage[dvipdfmx]{hyperref}
\usepackage{url}
\hypersetup{
    colorlinks=true,
    citecolor=cyan,
}
\newcommand*{\D}{{\rm d}}
\newcommand*{\mpl}{M_{\rm Pl}}
\begin{document}

\title{Generalized multi-Galileons, covariantized new terms, and the no-go theorem for non-singular cosmologies}

\author{Shingo~Akama}
\email[Email: ]{s.akama"at"rikkyo.ac.jp}
\affiliation{Department of Physics, Rikkyo University, Toshima, Tokyo 171-8501, Japan
}
\author{Tsutomu~Kobayashi}
\email[Email: ]{tsutomu"at"rikkyo.ac.jp}
\affiliation{Department of Physics, Rikkyo University, Toshima, Tokyo 171-8501, Japan
}

\begin{abstract}
It has been pointed out that non-singular cosmological solutions
in second-order scalar-tensor theories
generically suffer from gradient instabilities.
We extend this no-go result to second-order gravitational theories
with an arbitrary number of interacting scalar fields.
Our proof follows directly from the action of generalized multi-Galileons,
and thus is different from
and complementary to that based on the effective field theory approach.
Several new terms for generalized multi-Galileons on a flat background were proposed recently.
We find a covariant completion of them and confirm that
they do not participate in the no-go argument.
\end{abstract}


\preprint{RUP-17-1}
\maketitle

\section{Introduction}

Inflation~\cite{Guth:1980zm,Starobinsky:1980te,Sato:1980yn} is
an attractive scenario because it gives a natural resolution of
the horizon and flatness problems in standard Big Bang cosmology
and accounts for the origin of density perturbations that
are consistent with observations such as CMB.
However, there are criticisms that even inflation cannot resolve
the initial singularity~\cite{Borde:1996pt}
and the trans-Planckian problem for cosmological perturbations~\cite{Martin:2000xs}.
Alternative scenarios such as bounces and Galilean Genesis
have therefore been explored by a number of authors (see, e.g., Ref.~\cite{Battefeld:2014uga} for a review).

To avoid the initial singularity, there must be a period in which
the Hubble parameter $H$ is an increasing function of time.
This indicates a violation of the null energy condition (NEC), possibly causing
some kind of instability. It is easy to show that
NEC-violating cosmological solutions are indeed unstable
if the Universe is filled with a usual scalar field or a perfect fluid.
However, this is not the case if the underlying Lagrangian depends on
second derivatives of a scalar field~\cite{Rubakov:2014jja},
and one can construct explicitly
a stable cosmological phase in which the NEC is violated
in the Galileon-type scalar-field theory~\cite{Creminelli:2010ba,Deffayet:2010qz,Kobayashi:2010cm}.

Nevertheless, this does not mean that such non-singular cosmological solutions
are stable at all times in the entire history; it has been known
that gradient instabilities occur at some moment in many concrete
examples (see, e.g.,
Refs.~\cite{Cai:2012va,Koehn:2013upa,Battarra:2014tga,Qiu:2015nha,Wan:2015hya,Pirtskhalava:2014esa,Kobayashi:2015gga}),
and in some cases the instabilities show up even in the far future
after the NEC violating stage~\cite{Qiu:2011cy,Easson:2011zy,Ijjas:2016tpn}.
Recently, it was shown that this is
a generic nature of non-singular cosmological solutions in the Horndeski/generalized Galileon
theory~\cite{Horndeski:1974wa,Deffayet:2011gz,Kobayashi:2011nu},
i.e., in the most general scalar-tensor theory having second-order field equations,
provided that graviton geodesics are complete~\cite{Libanov:2016kfc,Kobayashi:2016xpl,Creminelli:2016zwa}.

As the no-go result is obtained in the single-field Horndeski theory,
one could evade this
by considering theories with multiple scalar fields or
higher derivative theories beyond Horndeski.
The latter way is indeed successful within
the Gleyzes-Langlois-Piazza-Vernizzi scalar-tensor theory~\cite{Gleyzes:2014dya,Gleyzes:2014qga,Zumalacarregui:2013pma},
as pointed out in Refs.~\cite{Cai:2016thi,Creminelli:2016zwa} based on the effective field theory (EFT) of
cosmological perturbations~\cite{Cheung:2007st}.
Gradient instabilities can also be cured
if higher spatial derivative terms arise in the action for curvature
perturbations~\cite{Creminelli:2006xe,Pirtskhalava:2014esa,Kobayashi:2015gga}.
This occurs in a more general framework~\cite{Gao:2014soa,Gao:2014fra}
than~\cite{Gleyzes:2014dya}
including Ho\v{r}ava gravity~\cite{Horava:2009uw}.
In some cases it is possible, even without such general frameworks, that
the strong coupling scale cuts off the instabilities~\cite{Koehn:2015vvy}.

The purpose of the present paper is to show that,
in contrast to the case of the higher derivative extension,
the no-go theorem for non-singular cosmologies still holds
in general multi-scalar-tensor theories of gravity.
In a subclass of the generalized multi-Galileon theory~\cite{Padilla:2012dx},
the same conclusion as in the single-field case was obtained in~\cite{Kolevatov:2016ppi}.
It was found in~\cite{Creminelli:2016zwa} that
the no-go theorem can also be extended to
the EFT of multi-field models
in which a shift symmetry is assumed for the entropy mode~\cite{Senatore:2010wk}.
(See Ref.~\cite{Noumi:2012vr} for the EFT of multi-field inflation without the shift symmetry.)
In this paper, we provide a new proof
which follows directly from the full action of the generalized multi-Galileon theory.

This paper is organized as follows.
In the next section, we give a brief review on the generalized multi-Galileon theory
and extend the proof of the no-go theorem for non-singular cosmologies
to multi-field models. Recently,
several new terms were found that are not included in the generalized multi-Galileon theory
but still yield second-order field equations~\cite{Allys:2016hfl}.
To keep the proof as general as possible,
we show in Sec.~III that the main result is not changed by the addition of these new terms.
In doing so, we find a covariant completion of the flat-space action of Ref.~\cite{Allys:2016hfl}.
In Sec.~IV we give a comment on the (in)completeness of graviton geodesics
viewed from the original (non-Einstein) frame.
We draw our conclusions in Sec.~V.



\section{No-go theorem in generalized multi-Galileon theory}

\subsection{Generalized multi-Galileon theory}

The most general single-scalar-tensor theory whose field equations are of
second order is given by the Horndeski action~\cite{Horndeski:1974wa}.
To begin with, let us review briefly how the same theory was rediscovered
in a different way starting from the Galileon theory.
The Galileon theory is a scalar-field theory on a fixed Minkowski background
having the Galilean shift symmetry, $\partial_\mu\phi\to\partial_\mu\phi + b_\mu$,
and second-order field equations~\cite{Nicolis:2008in}.
To make the metric dynamical and consider an arbitrary spacetime,
one can covariantize the Galileon theory by replacing $\partial_\mu$ with $\nabla_\mu$,
but this procedure induces higher derivative terms in the field equations
due to the noncommutativity of the covariant derivative.
However, the resulting higher derivative terms can be removed
by introducing non-minimal derivative coupling to the curvature.
The covariant multi-Galileon theory is thus obtained~\cite{Deffayet:2009wt}.
Now the Galilean shift symmetry is lost and
what is more important is the second-order nature of the field equations,
as it guarantees the absence of Ostrogradski instabilities.
One can further generalize the covariant Galileon theory by promoting
$X:=-g^{\mu\nu}\partial_\mu\phi\partial_\nu\phi/2$ in the action to
arbitrary functions $\phi$ and $X$ while retaining the second-order field equations~\cite{Deffayet:2011gz}.
This yields the Lagrangian
\begin{align}
{\cal L}&=G_2(X,\phi)-G_3(X,\phi)\Box\phi+G_4(X,\phi )R
\notag \\ & \quad
+\frac{\partial G_4}{\partial X}\left[(\Box\phi)^2-(\nabla_\mu\nabla_\nu\phi)^2 \right]
+G_5(X,\phi)G^{\mu\nu}\nabla_\mu\nabla_\nu\phi
\notag \\ & \quad
-\frac{1}{6}\frac{\partial G_5}{\partial X}\left[(\Box\phi)^3-3\Box\phi(\nabla_\mu\nabla_\nu\phi)^2 
+2(\nabla_\mu\nabla_\nu\phi)^3\right],\label{actionHor}
\end{align}
where $R$ is the Ricci scalar and $G_{\mu\nu}$ is the Einstein tensor.
Interestingly, it can be shown that this Lagrangian is equivalent to
the one obtained by Horndeski in an apparently different form~\cite{Kobayashi:2011nu},
and therefore is the most general one having second-order field equations.

The multi-field generalization can proceed in the following way.
In Refs.~\cite{Deffayet:2010zh,Padilla:2010de,Padilla:2010ir,Padilla:2010ir2,Trodden:2011xh,Sivanesan:2013tba},
the Galileons on
a fixed Minkowski background was
generalized to multi-field models,
whose action is
a functional of $N$ scalar fields $\phi^I$ ($I=1,\,2,\, ...,\,N$)
and their derivatives of order up to two.
Covariantizing the multi-Galileons and
introducing arbitrary functions of the scalar fields and their first derivatives
so that no higher derivative terms appear in the field equations,
one can arrive at the generalized multi-Galileon theory, the Lagrangian of which is given
in an analogous form to Eq.~(\ref{actionHor})
by~\cite{Padilla:2012dx}
\begin{align}
\mathcal{L}&=G_2(X^{IJ},{\phi}^K)-G_{3L}(X^{IJ},{\phi}^K)\Box{\phi}^L+G_4(X^{IJ},{\phi}^K)R\nonumber\\
&
\quad
+G_{4,\langle{IJ}\rangle}
\bigl(
{\Box{\phi}}^I{\Box{\phi}}^J-{\nabla}_{\mu}{\nabla}_{\nu}{\phi}^I{\nabla}_{\mu}{\nabla}_{\nu}{\phi}^J
\bigr)
\nonumber\\
&
\quad
+G_{5L}(X^{IJ},{\phi}^K)G^{\mu\nu}{\nabla}_{\mu}{\nabla}_{\nu}{\phi}^L
-\frac{1}{6}G_{5I,\langle{JK}\rangle}
\nonumber\\
&
\quad \quad
\times \bigl(\Box{\phi}^I\Box{\phi}^J\Box{\phi}^K
-3\Box{\phi}^{(I}{\nabla}_{\mu}{\nabla}_{\nu}{\phi}^J{\nabla}^{\mu}{\nabla}^{\nu}{\phi}^{K)}
\nonumber\\
&
\quad \quad
+2{\nabla}_{\mu}{\nabla}_{\nu}{\phi}^I{\nabla}^{\nu}{\nabla}^{\lambda}{\phi}^J{\nabla}_{\lambda}{\nabla}^{\mu}{\phi}^K
\bigr),\label{multi-G-L}
\end{align}
where
\begin{align}
X^{IJ}&:=-\frac{1}{2}g^{\mu\nu}{\partial}_{\mu}{\phi}^I{\partial}_{\nu}{\phi}^J,
\\
G_{,\langle IJ \rangle}
&:=\frac{1}{2}\left(\frac{\partial{G}}{\partial{X^{IJ}}}+\frac{\partial{G}}{\partial{X^{JI}}}\right).
\end{align}
In order for the field equations to be of second order,
it is required that
\begin{align}
& G_{3IJK}:=G_{3I,\langle JK\rangle},
&& G_{4IJKL}:=G_{4,\langle IJ \rangle,\langle KL\rangle},
\\
& G_{5IJK}:=G_{5I,\langle JK\rangle},
&& G_{5IJKLM}:=G_{4IJK,\langle LM \rangle},
\end{align}
are symmetric in all of their indices $I, \,J, \,...$.
In what follows we will write $G_{4,\langle IJ\rangle}$ as $G_{4IJ}$.
It is obvious that $G_{4IJ}=G_{4JI}$.

The multi-scalar-tensor theory described by the Lagrangian~(\ref{multi-G-L})
seems very general and includes
the earlier works~\cite{Damour:1992we, Horbatsch:2015bua}
and more recent ones~\cite{Kolevatov:2016ppi,Naruko:2015zze,Charmousis:2014zaa,Saridakis:2016ahq,Saridakis:2016mjd}
as specific cases. However, in contrast to the case of the single Galileon,
it is {\em not} the most general multi-scalar-tensor theory with second-order field equations.
Indeed, as demonstrated in~\cite{Kobayashi:2013ina},
the multi-DBI Galileon theory~\cite{RenauxPetel:2011uk}
is not included in the above one.
To date, no complete multi-field generalization of the Horndeski action has been known.
Taking the same approach as Horndeski did
rather than starting from the multi-Galileon theory,
the authors of Ref.~\cite{Ohashi:2015fma} obtained the most general second-order
field equations of {\em bi}-scalar-tensor theories, but
deducing the corresponding action and
extending the bi-scalar result to the case of more than two scalars have not been successful so far.
We will come back to this issue in the next section
in light of the recent result reported in~\cite{Allys:2016hfl}.

Although the generalized multi-Galileon theory is thus not the most general one,
it is definitely quite general and so we choose to
use the Lagrangian~(\ref{multi-G-L}).
This is one of the best things one can do at this stage to
draw some general conclusions on the cosmology of multiple interacting scalar fields,
and is considered as complementary to the
approach based on the effective field theory of multifield inflation~\cite{Creminelli:2016zwa}.

\subsection{Stability of a non-singular universe in generalized multi-Galileon theory}

We now show that the no-go theorem in~\cite{Kobayashi:2016xpl}
can be extended to the case of the generalized multi-Galileon theory.

The quadratic actions for
perturbations around a flat Friedmann background have been calculated in~\cite{Kobayashi:2013ina}.
For tensor perturbations $h_{ij}(t,\Vec{x})$ we have
\begin{align}
S_h^{(2)}=
\frac{1}{8} \int \D t\D^3x \,
a^3\left[
{\mathcal G}_T\dot{h}_{ij}^2-\frac{{\mathcal F}_T}{a^2}
(\Vec{\nabla}h_{ij})^2
\right],\label{ac2tens}
\end{align}
where
\begin{align}
{\mathcal G}_T:=2 \left[ G_4-2X^{IJ}G_{4IJ}-X^{IJ}(H\dot{\phi}^KG_{5IJK}-G_{5I,J}) \right]
\end{align}
and
\begin{align}
{\mathcal F}_T:=2 \left[ G_4-X^{IJ}(\ddot{\phi}^KG_{5IJK}+G_{5I,J}) \right].
\end{align}
Here we defined
$G_{,I}:=\partial{G}/\partial{\phi}^I$.
Stability requires
\begin{align}
{\cal G}_T>0,\quad {\cal F}_T>0,
\end{align}
at any moment in the whole cosmological history.

To study scalar perturbations in multi-field models, it is convenient to use the spatially flat gauge.
The quadratic action for scalar perturbations
is of the form~\cite{Kobayashi:2013ina}
\begin{align}
S_Q^{(2)}
=\frac{1}{2}{\int}\D t\D^3x
a^3 &
\biggl[
{\cal K}_{IJ}\dot Q^I \dot Q^J-\frac{1}{a^2}{\cal D}_{IJ}\Vec{\nabla}Q^I\cdot\Vec{\nabla}Q^J
\notag \\
&
-{\cal M}_{IJ}Q^IQ^J+2\Omega_{IJ}Q^I\dot Q^J
\biggr],
\end{align}
where $Q^I$'s are the perturbations of the scalar fields defined by
\begin{equation}
\phi^I=\bar{\phi}^I(t)+Q^I(t,\vec{x}).
\end{equation}
The explicit expressions for the matrices
${\cal K}_{IJ}$, ${\cal M}_{IJ}$, and $\Omega_{IJ}$
can be found in~\cite{Kobayashi:2013ina},
but are not necessary for the following discussion.
Since gradient instabilities manifest most significantly at high frequencies, only
the structure of the matrix ${\cal D}_{IJ}$ is crucial to our no-go argument.
We will use the fact that ${\cal D}_{IJ}$ is given by~\cite{Kobayashi:2013ina}
\begin{align}
{\cal D}_{IJ}={\cal C}_{IJ}-\frac{{\cal J}_{(I}{\cal B}_{J)}}{\Theta}
+\frac{1}{a}\frac{\D}{\D t}\left(
\frac{a{\cal B}_I{\cal B}_J}{2\Theta}
\right),\label{DIJrelation}
\end{align}
where ${\cal C}_{IJ}$ is the matrix satisfying the identity
\begin{align}
{\cal C}_{IJ}X^{IJ}=2H\left(\dot{\cal G}_T+H{\cal G}_T\right)
-\dot\Theta-H\Theta-H^2{\cal F}_T,\label{XIJC}
\end{align}
with
\begin{align}
\Theta &:= 
-\dot\phi^IX^{JK}G_{3IJK} + 2HG_4
\notag \\ & \quad
-8HX^{IJ}\left(G_{4IJ}+X^{KL}G_{4IJKL}\right)
\notag \\ & \quad
+2\dot\phi^IX^{JK}G_{4IJ, K}+\dot\phi^IG_{4,I}
\notag \\ & \quad
-H^2\dot\phi^IX^{JK}\left(5G_{5IJK}+2X^{LM}G_{5IJKLM}\right)
\notag \\ & \quad
+2HX^{IJ}\left(
3G_{5I,J}+2X^{KL}G_{5IJK,L}
\right).\label{deftheta}
\end{align}
The explicit expressions for
${\cal J}_I$ and ${\cal B}_I$ in Eq.~(\ref{DIJrelation})
are also unimportant, but
we will use the equation~\cite{Kobayashi:2013ina}
\begin{align}
\dot\phi^I{\cal J}_I+\ddot\phi^I{\cal B}_I+2\dot H{\cal G}_T=0.\label{idFRD}
\end{align}
This follows from the background equations,
and corresponds in the minimally coupled single-field case to
the familiar equation
\begin{align}
\dot\phi^2+2\mpl^2\dot H = 0.
\end{align}

It is required for the stability of the scalar sector that
the matrices ${\boldsymbol {\cal K}}=({\cal K}_{IJ})$ and ${\boldsymbol {\cal D}}=({\cal D}_{IJ})$ must be
positive definite. Hence, a non-singular cosmological solution
is free from gradient instabilities if, for
every non-zero column vector ${\boldsymbol v}$,
\begin{align}
{\boldsymbol v}^{{\rm T}}{\boldsymbol {\cal D}}{\boldsymbol v}>0,\label{gradsta}
\end{align}
where ${\boldsymbol v}^{{\rm T}}$ is the transpose of ${\boldsymbol v}$.
Now, let ${\boldsymbol v}$ be
\begin{align}
{\boldsymbol v}=\left(
    \begin{array}{c}
      \dot\phi^1 \\
      \dot\phi^2 \\
      \vdots \\
      \dot\phi^N
    \end{array}
  \right).
\end{align}
Then, Eq.~(\ref{gradsta}) reads
\begin{align}
{\boldsymbol v}^{{\rm T}}{\boldsymbol {\cal D}}{\boldsymbol v}=2X^{IJ}{\cal D}_{IJ}>0.
\end{align}
Using Eqs.~(\ref{DIJrelation}),~(\ref{XIJC}), and~(\ref{idFRD})
and doing some manipulation, one finds
\begin{align}
X^{IJ}{\cal D}_{IJ} = H^2\left(\frac{1}{a}\frac{\D\xi}{\D t}-{\cal F}_T\right)>0,\label{ineq1}
\end{align}
where
\begin{align}
\xi:=\frac{a{\mathcal G}_T^2}{\Theta}.
\end{align}

The remaining part of the proof is parallel to that in the Horndeski case~\cite{Kobayashi:2016xpl},
because the structure of the inequality~(\ref{ineq1}) is identical to the single-field counterpart.
In a non-singular universe, $\Theta$ never diverges
because it is composed of $H$ and $\phi^I$ as given in Eq.~(\ref{deftheta})
and we require that the functions $G_2$, $G_{3I}$, ... in the underlying Lagrangian
remain finite in the entire cosmological history.\footnote{Our postulate on this point is
different from that adopted in Ref.~\cite{Ijjas:2016tpn}, in which {\em singular} functions
are introduced in the underlying Lagrangian to obtain non-singular cosmological solutions.}
We also have $a{\cal G}_T^2>0$ which comes from the stability of the tensor perturbations.\footnote{Our
postulate on this point is different from that adopted in Ref.~\cite{Ijjas:2016wtc},
in which all the coefficients in the quadratic action for cosmological perturbations
vanish at the same moment.}
Therefore, $\xi$ cannot cross zero.
From Eq.~(\ref{ineq1}) we have
\begin{align}
\frac{\D \xi}{\D t}>a{\cal F}_T>0,\label{ineq2}
\end{align}
indicating that $\xi$ is a monotonically increasing function of $t$.
Integrating Eq.~(\ref{ineq2}) from some $t_{\rm i}$ to $t_{\rm f}$,
we obtain
\begin{align}
\xi(t_{\rm f})-\xi(t_{\rm i}) > \int_{t_{\rm i}}^{t_{\rm f}} a{\cal F}_T\D t'.\label{ineq3}
\end{align}
(We admit that $\xi$ diverges at some $t_\ast$ where $\Theta=0$ occurs.
In this case, $t_{\rm i}$ and $t_{\rm f}$ are taken to
be such that $t_{\rm i}<t_{\rm f}<t_\ast$ or $t_\ast<t_{\rm i}< t_{\rm f}$.)
If $\lim_{t\to-\infty}\xi=\,$const, we take $t_{\rm i}\to-\infty$ in Eq.~(\ref{ineq3})
and obtain
\begin{align}
\int_{-\infty}^{t_{\rm f}}a{\cal F}_T\D t'<\xi(t_{\rm f})-\xi(-\infty)<\infty.
\end{align}
Similarly, if $\lim_{t\to\infty}\xi=\,$const then we take $t_{\rm f}\to\infty$
to get
\begin{align}
\int^{\infty}_{t_{\rm i}}a{\cal F}_T\D t'<\xi(\infty)-\xi(t_{\rm i})<\infty.
\end{align}
Thus, we conclude that a non-singular cosmological solution in the generalized multi-Galileon theory
is stable in the entire history provided that either
\begin{align}
\int_{-\infty}^ta{\cal F}_T\D t'\quad
{\rm or} \quad
\int_t^\infty a{\cal F}_T\D t'\label{convint}
\end{align}
is convergent. (If $\Theta = 0$ occurs, both of the above integrals must be convergent.)
As is argued in Refs.~\cite{Creminelli:2016zwa,Cai:2016thi} and also in Sec.~IV of the present paper,
the convergence of the above integrals signals some kind of pathology in
the tensor perturbations. If one prefers to avoid this pathology, all non-singular cosmological solutions
in the generalized multi-Galileon theory are inevitably plagued with gradient instabilities.

One might expect naively that, in the presence of multiple interacting scalar fields,
a dominant field can transfer its energy to another field or matter before
the instability of the former shows up, and thus the instability
can be eliminated.
We have shown that this is not the case in the generalized multi-Galileon theory.

The same conclusion was reached
using the EFT of multi-field cosmologies, in which a shift symmetry
is assumed for the entropy mode~\cite{Creminelli:2016zwa}.
Our proof is different from, and complementary to, that based on the EFT.
The EFT approach amounts to writing all the terms allowed by symmetry,
which leads to the theory of cosmological perturbations on a given background.
Therefore, the adiabatic and entropy modes are decomposed
by construction in the EFT.
In contrast, our guiding principle is the second-order nature of the field equations,
and so we start from the general action of second-order multiple scalar-tensor theories
that governs the perturbation evolution as well as the background dynamics.
It should be noticed that we have not performed the adiabatic/entropy decomposition,
as it is unnecessary for our no-go argument.
Although the relation between the second-order theory and
the EFT of cosmological perturbations has been clarified
in the single-field case~\cite{Gleyzes:2013ooa}, to date, it is not obvious how
the EFT of multi-field cosmology is related to
the generalized multi-Galileon theory.

\section{Covariantized new terms for multi-Galileon theory}

Very recently, the author of Ref.~\cite{Allys:2016hfl}
proposed new terms for scalar multi-Galileon theory
that are not included in the existing multi-Galileon Lagrangian
but give rise to a second-order field equation.
The Lagrangians for these ``extended'' multi-Galileons are given by~\cite{Deffayet:2010zh,Allys:2016hfl} 
\begin{align}
{\cal L}_{{\rm ext}1}&=
A_{[IJ][KL]M}\delta_{\nu_1\nu_2\nu_3}^{\mu_1\mu_2\mu_3}
\partial_{\mu_1}\phi^I\partial_{\mu_2}\phi^J\partial^{\nu_1}\phi^K\partial^{\nu_2}\phi^L
\notag \\ &\quad \times
\partial_{\mu_3}\partial^{\nu_3}\phi^M,\label{fext1}
\\
{\cal L}_{{\rm ext}2}&=
A_{[IJ][KL](MN) }
\delta_{\nu_1\nu_2\nu_3\nu_4}^{\mu_1\mu_2\mu_3\mu_4}
\partial_{\mu_1}\phi^I\partial_{\mu_2}\phi^J
\notag \\ &\quad \times
\partial^{\nu_1}\phi^K\partial^{\nu_2}\phi^L
\partial_{\mu_3}\partial^{\nu_3}\phi^M\partial_{\mu_4}\partial^{\nu_4}\phi^N,\label{fext2}
\\
{\cal L}_{{\rm ext}3}&=
A_{[IJK][LMN]O }
\delta_{\nu_1\nu_2\nu_3\nu_4}^{\mu_1\mu_2\mu_3\mu_4}
\partial_{\mu_1}\phi^I\partial_{\mu_2}\phi^J\partial_{\mu_3}\phi^K
\notag \\ &\quad \times
\partial^{\nu_1}\phi^L\partial^{\nu_2}\phi^M\partial^{\nu_3}\phi^N
\partial_{\mu_4}\partial^{\nu_4}\phi^O,\label{fext3}
\end{align}
where the coefficients $A_{[IJ][KL]M},\, ...$ are arbitrary functions of
$\phi^I$ and $X^{IJ}$. These coefficients are
antisymmetric in indices inside $[~]$ and symmetric in indices inside $(~)$.
In order for the field equations to be of second order,
we require that
\begin{align}
&A_{[IJ][KL]\underline{M,\langle NO\rangle}},
\quad
A_{[IJ][KL]\underline{(MN),\langle OP\rangle}},
\notag \\ &
A_{[IJK][LMN]\underline{O,\langle PQ\rangle}},
\end{align}
are symmetric in underlined indices.

The Lagrangians (\ref{fext1})--(\ref{fext3})
are those for scalar fields on fixed Minkowski spacetime.
Let us explore a covariant completion of the above flat-space multi-scalar theory.
To make the metric dynamical, we first promote $\partial_\mu$ to $\nabla_\mu$.
It is easy to see that this procedure is sufficient for
${\cal L}_{{\rm ext}1}$ and ${\cal L}_{{\rm ext}3}$:
\begin{align}
{\cal L}_{{\rm ext}1}'&=
A_{[IJ][KL]M}\delta_{\nu_1\nu_2\nu_3}^{\mu_1\mu_2\mu_3}
\nabla_{\mu_1}\phi^I\nabla_{\mu_2}\phi^J\nabla^{\nu_1}\phi^K\nabla^{\nu_2}\phi^L
\notag \\ &\quad \times
\nabla_{\mu_3}\nabla^{\nu_3}\phi^M,\label{cext1}
\\
{\cal L}_{{\rm ext}3}'&=
A_{[IJK][LMN]O }
\delta_{\nu_1\nu_2\nu_3\nu_4}^{\mu_1\mu_2\mu_3\mu_4}
\nabla_{\mu_1}\phi^I\nabla_{\mu_2}\phi^J\nabla_{\mu_3}\phi^K
\notag \\ &\quad \times
\nabla^{\nu_1}\phi^L\nabla^{\nu_2}\phi^M\nabla^{\nu_3}\phi^N
\nabla_{\mu_4}\nabla^{\nu_4}\phi^O,\label{cext3}
\end{align}
have second-order equations of motion for the metric and scalar fields.
However, the simple covariantization of ${\cal L}_{{\rm ext}2}$,
\begin{align}
{\cal L}_{{\rm cext}2}&=
A_{[IJ][KL](MN) }
\delta_{\nu_1\nu_2\nu_3\nu_4}^{\mu_1\mu_2\mu_3\mu_4}
\nabla_{\mu_1}\phi^I\nabla_{\mu_2}\phi^J
\notag \\ &\quad \times
\nabla^{\nu_1}\phi^K\nabla^{\nu_2}\phi^L
\nabla_{\mu_3}\nabla^{\nu_3}\phi^M\nabla_{\mu_4}\nabla^{\nu_4}\phi^N,
\end{align}
yields higher derivative terms in the field equations.
To cancel such terms, we add a counter term, i.e.,
a coupling to the curvature tensor ${\cal L}_{{\rm curv}2}$.
It turns out that the appropriate Lagrangian is the following:
\begin{align}
{\cal L}_{{\rm curv}2}&=B_{[IJ][KL]}\delta_{\nu_1\nu_2\nu_3\nu_4}^{\mu_1\mu_2\mu_3\mu_4}
\notag \\ &
\quad \times
R^{\nu_3\nu_4}_{~~~~~\mu_3\mu_4}
\nabla_{\mu_1}\phi^I\nabla_{\mu_2}\phi^J\nabla^{\nu_1}\phi^K\nabla^{\nu_2}\phi^L,
\end{align}
where
\begin{align}
B_{[IJ][KL],\langle MN\rangle }=\frac{1}{2}A_{[IJ][KL](MN)}
\end{align}
must be imposed.
Thus, we find that the covariant completion of ${\cal L}_{{\rm ext}2}$ is given by
\begin{align}
{\cal L}_{{\rm ext}2}'&={\cal L}_{{\rm curv}2}+{\cal L}_{{\rm cext}2}
\end{align}
where $A_{[IJ][KL](MN)}=2B_{[IJ][KL],\langle MN\rangle }$ and
\begin{align}
B_{[IJ][KL]\underline{MNOP}}:=B_{[IJ][KL],\langle MN\rangle,\langle OP\rangle }
\end{align}
is symmetric in underlined indices.

One can check that
the multi-DBI Galileon theory at leading order in the $X^{IJ}$ expansion~\cite{Kobayashi:2013ina}
is obtained by taking 
\begin{align}
B_{[IJ][KL]}= {\rm const}\times \left(\delta_{IK}\delta_{JL}-\delta_{IL}\delta_{JK}\right),
\end{align}
though it seems extremely difficult to see explicitly that
the complete Lagrangian for the multi-DBI Galileons~\cite{RenauxPetel:2011uk}
can be reproduced by choosing appropriately the functions in the above Lagrangians.

Now the question is how the additional terms
\begin{align}
{\cal L}_{\rm ext}:={\cal L}_{{\rm ext}1}'+{\cal L}_{{\rm ext}2}'+{\cal L}_{{\rm ext}3}'
\end{align}
change the stability of cosmological solutions.
Obviously, ${\cal L}_{\rm ext}$ does not change the background equations due to antisymmetry.
We see that, in the quadratic actions for scalar and tensor perturbations,
only the ${\cal C}_{IJ}$ coefficients are modified as follows:
\begin{align}
{\cal C}_{IJ}\to{\cal C}_{IJ}+{\cal C}_{IJ}^{\rm ext},
\end{align}
with
\begin{align}
{\cal C}_{IJ}^{\rm ext}&:=32H\bigl(
-A_{[IK][JL]M}X^{KL}\dot\phi^M
+2HB_{[IK][JL]}X^{KL}
\notag \\ &\quad\qquad\quad
+4HB_{[IK][JL],\langle MN\rangle}X^{KL}X^{MN}
\bigr),
\end{align}
and no other terms are affected by the addition of ${\cal L}_{\rm ext}$.
Since $X^{IJ}{\cal C}_{IJ}^{\rm ext}=0$ due to antisymmetry,
$X^{IJ}{\cal D}_{IJ}$ remains the same even if one adds ${\cal L}_{\rm ext}$:
\begin{align}
X^{IJ}{\cal D}_{IJ}\to X^{IJ}{\cal D}_{IJ}.
\end{align}
Therefore, the new terms proposed in Ref.~\cite{Allys:2016hfl}
do not change the no-go argument.

The new term ${\cal L}_{{\rm ext}}$ vanishes for the homogeneous background,
which implies that ${\cal L}_{{\rm ext}}$ contributes only to the
entropy modes at the level of perturbations.
This is consistent with the result of~\cite{Creminelli:2016zwa},
where it can be seen using the EFT that the instability occurs in the adiabatic direction.

\section{Graviton geodesics}

We have thus seen that within the multi-field extension of the generalized Galileons,
non-singular cosmological solutions are possible only if
either integral
in Eq.~(\ref{convint}) is convergent,
as in the single-field Horndeski case.
In Ref.~\cite{Kobayashi:2016xpl}, this fact was noticed
and a numerical example of a non-singular cosmological solution
with the convergent integral
was obtained for the first time in the single-field context.
Later, the authors of Ref.~\cite{Ijjas:2016vtq} followed Ref.~\cite{Kobayashi:2016xpl} and presented another example.

One can move to the ``Einstein frame'' for tensor perturbations
from the original frame~(\ref{ac2tens})
by performing
a disformal transformation~\cite{Creminelli:2014wna}.
This is because one has two independent functions of $t$ in performing a disformal transformation
which can be fitted to make ${\cal F}_T$ and ${\cal G}_T$ into their standard forms:
${\cal F}_T\to\mpl^2$, ${\cal G}_T\to \mpl^2$.
It is clearly explained in Ref.~\cite{Creminelli:2016zwa} that because
gravitons propagate along null geodesics in the Einstein frame and
the integral
\begin{align}
\int a{\cal F}_T\D t \label{int2grav}
\end{align}
is nothing but the affine parameter of the null geodesics in the Einstein frame,
the convergent integral~(\ref{convint}) implies
past (future) incompleteness of graviton geodesics (see also Ref.~\cite{Cai:2016thi}).
This may signal some kind of pathology in the tensor perturbations,
though it is not obvious whether the incompleteness of null geodesics
in a disformally related frame causes actual problems.

Let us rephrase this potential pathology of gravitons
without invoking the disformal transformation.
The equation of motion for the tensor perturbation $h_{ij}$
derived from the action~(\ref{ac2tens}) can be written in the form
\begin{align}
Z^{\mu\nu}{\cal D}_\mu{\cal D}_\nu h_{ij}=0,\label{effgeoh}
\end{align}
where
\begin{align}
Z_{\mu\nu}\D x^\mu\D x^\nu =-\frac{{\cal F}_T^{3/2}}{{\cal G}_T^{1/2}}\D t^2
+a^2\left({\cal F}_T{\cal G}_T\right)^{1/2}\delta_{ij}\D x^i\D x^j,
\end{align}
and ${\cal D}_\mu$ is the covariant derivative associated with
the ``metric'' $Z_{\mu\nu}$.
Equation~(\ref{effgeoh}) shows that
graviton paths can be interpreted as null geodesics in the effective geometry
defined by $Z_{\mu\nu}$.
It turns out that
the affine parameter $\lambda$ of null geodesics in the metric $Z_{\mu\nu}$
is given by $\D\lambda = a{\cal F}_T\D t$.
Therefore, the incompleteness of graviton geodesics can be made manifest
even without working in the Einstein frame.

\section{Summary}

In this paper,
we have shown that
all non-singular cosmological solutions are plagued with gradient instabilities
in the multi-field generalization of scalar-tensor theories,
if the graviton geodesic completeness is required.
This extends the recent no-go arguments of Refs.~\cite{Libanov:2016kfc,Kobayashi:2016xpl,Kolevatov:2016ppi}.
We have given a direct proof using
the generalized multi-Galileon action, so that
our proof is different from and complementary to
that obtained from the effective field theory of cosmological fluctuations~\cite{Creminelli:2016zwa}.
Several new terms for multi-Galileons on a flat background
were found recently~\cite{Allys:2016hfl}. We have covariantized these terms
and shown that the inclusion of them does not change the no-go result.

\acknowledgments
We thank Yuji Akita, Norihiro Tanahashi, Masahide Yamaguchi,
and Shuichiro Yokoyama for helpful discussions.
This work was supported in part
by the MEXT-Supported Program for the Strategic Research Foundation at Private Universities, 2014-2017,
and by the JSPS Grants-in-Aid for
Scientific Research No.~16H01102 and No.~16K17707 (T.K.).





\end{document}